\documentclass{mnras}
\usepackage{graphicx}
\usepackage{times}
\usepackage[dvipsnames]{xcolor}
\usepackage{color}

\usepackage{cleveref}
\hypersetup{colorlinks={true},linkcolor={blue},citecolor=green}

\usepackage{bm}
\usepackage{amssymb}
\DeclareMathAlphabet{\mathpzc}{OT1}{pzc}{m}{it}

\usepackage{pgfplots}
\usetikzlibrary{patterns}

\catcode`\@=11
\newcommand\gsim{\ifmmode{\mathrel{\mathpalette\@versim>}}
    \else{$\mathrel{\mathpalette\@versim>}$}\fi}
\newcommand\lsim{\ifmmode{\mathrel{\mathpalette\@versim<}}
    \else{$\mathrel{\mathpalette\@versim<}$}\fi}
\catcode`\@=12

\newcommand\rhos{\rho_*}
\newcommand\rhog{\rho_{\rm g}}

\newcommand\rhoDM{\rho_{\rm DM}}
\newcommand\rhon{\rho_{\rm n}}

\newcommand\rhoNFW{\rho_{\rm NFW}}
\newcommand\Mmon{\mathcal{M}}

\newcommand\Hcsi{{\cal H}}
\newcommand\Litwo{{\rm Li}_2}
\newcommand\Fiso{{\mathcal{F}}_{\rm iso}}
\newcommand\Frad{{\mathcal{F}}_{\rm rad}}
\newcommand\vc{v_{\rm c}}
\newcommand\atan{{\rm arctg}}
\newcommand\K{\mathpzc K}

\newcommand\rs{r_*}
\newcommand\rg{r_{\rm g}}

\newcommand\reff{R_{\rm e}}
\newcommand\ra{r_{\rm a}}
\newcommand\sa{s_{\rm a}}
\newcommand\rt{r_{\rm t}}

\newcommand\Phig{\Phi_{\rm g}}
\newcommand\Psig{\Psi_{\rm g}}
\newcommand\Phibh{\Phi_{\rm BH}}

\newcommand\Phidm{\Phi_{\rm DM}}

\newcommand\Phis{\Phi_*}
\newcommand\Psis{\Psi_*}

\newcommand\PsiT{\Psi_{\rm T}}
\newcommand\Psin{\Psi_{\rm n}}

\newcommand\Ms{M_*}

\newcommand\Mbh{M_{\rm {BH}}}
\newcommand\Mg{M_{\rm g}}
\newcommand\MT{M_{\rm T}}

\newcommand\Mps{M_{{\rm p}*}}
\newcommand\MpDM{M_{\rm pDM}}
\newcommand\Mpg{M_{\rm pg}}
\newcommand\MR{{\cal R}}
\newcommand\MRN{{\cal R}_{\rm NFW}}
\newcommand\Rm{{\mathcal{R}}_{\rm m}}
\newcommand\Rmon{{\mathcal{R}}_{\rm mon}}

\newcommand\gs{g_*}
\newcommand\fstar{f_*}
\newcommand\ggal{g_{\rm g}}
\newcommand\fgal{f_{\rm g}}

\newcommand\qt{\tilde q}

\newcommand\Er{\mathcal{E}}

\newcommand\It{\mathcal{I}_{\rm g}}
\newcommand\At{\mathcal{A}_{\rm g}}
\newcommand\Ibh{\mathcal{I}_{\rm BH}}
\newcommand\Abh{\mathcal{A}_{\rm BH}}

\newcommand\srad{\sigma_{\rm r}}
\newcommand\stan{\sigma_{\rm t}}

\newcommand\Ks{K_*}
\newcommand\Krad{K_{\rm *r}}
\newcommand\Ktan{K_{\rm *t}}
\newcommand\Ws{W_*}
\newcommand\Wss{W_{**}}
\newcommand\Wg{W_{\rm *g}}
\newcommand\Wbh{W_{\rm *BH}}
\newcommand\Wdm{W_{\rm *DM}}
\newcommand\Us{U_*}
\newcommand\Uss{U_{**}}
\newcommand\Ug{U_{\rm *g}}
\newcommand\Ubh{U_{\rm *BH}}
\newcommand\Udm{U_{\rm *DM}}
\newcommand\Bg{B_{*\rm g}}

\newcommand\Sigs{\Sigma_*}
\newcommand\Sigg{\Sigma_{\rm g}}

\newcommand\sigp{\sigma_{\rm p}}

\newcommand\sam{\sa^{-}}
\newcommand\sap{\sa^{+}}
\newcommand\cond{{\cal C}}
\newcommand\condm{\cond_{-}}
\newcommand\condp{\cond_{+}}
\newcommand\MD{M_{\rm {DM}}}

\newcommand\csih{\xi_{\rm NFW}}

\title[Two-component galaxy models with a central BH]{A new class of galaxy models with a central BH - I. The spherical case}

   \author[L. Ciotti, A. Mancino, S. Pellegrini]
          {Luca Ciotti$^1$, Antonio Mancino$^1$, Silvia Pellegrini$^1$
\\$^1$Department of Physics and Astronomy, 
      University of Bologna, via Gobetti 93/3, 40129 Bologna, Italy}

\date{Accepted, September 17, 2019}
\pubyear{...}

\begin{document} 
\maketitle

\begin{abstract} 

  \noindent
  The dynamical properties of spherically symmetric galaxy models,
  where a Jaffe (1983) stellar density profile is embedded in a total
  mass density decreasing as $r^{-3}$ at large radii, are presented.
  The orbital structure of the stellar component is described by the
  Osipkov--Merritt anisotropy; the dark matter halo is isotropic, and
  a black hole is added at the center of the galaxy.  First, the
  conditions for a nowhere negative and monotonically decreasing dark
  matter halo density profile are derived; this profile can be made
  asymptotically coincident with a NFW profile at the center and at
  large radii.  Then the minimum value of the anisotropy radius for
  phase-space consistency is derived as a function of the galaxy
  parameters. The Jeans equations for the stellar component are solved
  analytically; the projected velocity dispersion at the center and at
  large radii is also obtained, for generic values of the anisotropy
  radius. Finally, analytical expressions for the terms entering the
  Virial Theorem are derived, and the fiducial anisotropy limit
  required to prevent the onset of Radial Orbit Instability is
  determined as a function of the galaxy parameters. The presented
  models, built following an approach already adopted in our previous
  works, can be a useful starting point for a more advanced modeling
  of the dynamics of elliptical galaxies, and can be easily
  implemented in numerical simulations requiring a realistic dynamical
  model of a galaxy.

\end{abstract}


\section{Introduction}

Spherically symmetric galaxy models, thanks to their simplicity, can
be useful in exploratory works in Stellar Dynamics (e.g., Bertin 2000,
Binney \& Tremaine 2008). A successful spherical model compensates its
geometric limitations with other features, such as the possibilities
to derive manageable analytical expressions for the most important
dynamical quantities, to easily include a dark matter (hereafter, DM)
halo with an adjustable density profile (or, alternatively, to specify
the total density profile), to model the dynamical effects of a
central black hole (hereafter, BH), and finally to control orbital
anisotropy. Once the model properties are controlled in the spherical
limit, then more sophisticated investigations, based on axisymmetric
or triaxial galaxy models, can be undertaken avoiding a large
exploration of the parameter space (e.g., Cappellari et al. 2007, van
den Bosch et al. 2008).

The analytical Jaffe (1983) density profile is a natural choice to
describe the stellar distribution of early-type galaxies in the
spherical approximation. It belongs to the family of the so-called
$\gamma$ models (Dehnen 1993, Tremaine et al. 1994), and resembles, in
projection, the de Vaucouleurs law (1984) $R^{1/4}$ with sufficient
accuracy (for most applications) over a large radial range.

Once the stellar profile of the model is considered acceptable, a
second request is the possibility to reproduce, with a minor effort,
the large scale properties of the {\it total} density profile
(e.g. Bertin et al. 1994, Rix et al. 1997, Gerhard et al. 2001; Treu
\& Koopmans 2002, 2004; Rusin et al. 2003; Rusin \& Kochanek 2005;
Koopmans et al. 2006; Gavazzi et al. 2007; Czoske et al. 2008; Dye et
al. 2008, Nipoti et al. 2008, see also Shankar et al. 2017). For
example, simple models with a flat rotation curve have in fact been
constructed (see, e.g., Kochaneck 1994, Naab \& Ostriker 2007, Ciotti
et al. 2009, hereafter CMZ09; see also the double power-law models of
Hiotelis 1994).

Finally, since supermassive BHs with a mass of the order of
$\Mbh\simeq 10^{-3}\Ms$ are routinely found at the center of the
stellar spheroids of total mass $\Ms$ (e.g., see Magorrian et
al. 1988, Kormendy \& Ho 2013), another feature of a useful spherical
model is the possibility to easily compute the dynamical properties of
the stellar component in presence of a central BH.

Following the arguments above, a family of models (hereafter, JJ
models) with a Jaffe profile for the stellar distribution, and a total
density profile described by another Jaffe law, so that the total mass
of JJ models kept finite, has already been proposed (Ciotti \& Ziaee
Lorzad 2018, hereafter CZ18).  For the stellar component JJ models
with a central BH, the Jeans equations with Osipkov-Merritt (Osipkov
1979, Merritt 1985, hereafter OM) radial anisotropy can be solved
analytically, and the projected velocity dispersion, at the center and
at large radii, can be expressed by means of simple
formulae. Moreover, for these models also the positivity of the
phase-space distribution function (hereafter, DF), the so-called
consistency, and the maximum amount of radial anisotropy allowable for
consistency, can be easily studied.

One interesting feature of the JJ models is that in the special {\it
  minumum halo} case, the DM profile, defined by the difference
between the total and the stellar profiles, behaves like $r^{-1}$ near
the center, similarly to the Navarro-Frenk-White profile (Navarro et
al. 1997, hereafter NFW); at large radii, instead, the DM profile
decreases as $r^{-4}$, at variance with the NFW profile that goes as
$r^{-3}$. The natural question left open is then if it is possible to
construct models with similar analytical properties of JJ models, but
with the additional property that the DM follows the $r^{-3}$ shape in
the external regions. In this paper we show that in fact this is
possible, and we call the resulting models ``J3'', to stress that the
stellar density is again a Jaffe model, while the DM decreases as
$r^{-3}$ in the external regions. In particular, we shall prove that
the DM halo, in minimum halo J3 models, can be made remarkably similar
to the NFW over the {\it whole} radial range.

The paper is organized as follows. In Sect. \ref{sect_models} the main
structural properties of the models are presented, and the conditions
required to have a nowhere negative and monotonically decreasing DM
halo density profile are derived; a discussion is also given of how
the DM component can be built in order to have the same asymptotical
behaviour, in the outer regions and near the center, as the NFW
profile. In Sect. \ref{sect_DF} we study the phase-space properties of
the models, and the minimum value of the anisotropy radius for
consistency is derived in terms of the galaxy parameters. In
Sect. \ref{sect_JEs} the analytical solution of the Jeans equations
with OM anisotropy is obtained, and the asymptotic trend of the
projected velocity dispersion profile at small and large radii is
given. Finally, in Sect. \ref{sec:Virial} the relevant global
quantities entering the Virial Theorem are explicitly calculated;
these are used for global energetic considerations, and to determine
the fiducial anisotropy limit required to prevent the onset of Radial
Orbit Instability as a function of the galaxy parameters. Section
\ref{sec:Conclusions} summarizes.

\section{The Models}\label{sect_models}

We name the proposed new family of models as ``J3'' models, to
indicate two-component models characterized by a {\it stellar} density
distribution $\rhos$ described by a Jaffe (1983) profile embedded in a
{\it total} density distribution $\rhog$ (stars plus DM) following a
$r^{-2}$ profile in the central regions and $r^{-3}$ at large
radii. The reasons for this choice will become clear in the following.

\subsection{Stellar distribution}\label{sec:Stellar_distr}

As in CZ18, the stellar component follows a Jaffe profile, with density and relative potential scales given by
\begin{equation}
\rhon\equiv\frac{\Ms}{4\pi\rs^3},\qquad \Psin\equiv\frac{G\Ms}{\rs},
\end{equation}
where $\Ms$ is the total stellar mass, and $\rs$ a scale length. In
these units, the stellar density-potential pair reads
\begin{equation}
\rhos(r)=\frac{\rhon}{s^2(1+s)^2}, \quad \Psis(r)=\Psin\ln\frac{1+s}{s},
\label{eq:rhos_psis}
\end{equation}
where $s\equiv r/\rs$ is the dimensionless radius, and in general we
indicate with $\Psi(r)\equiv -\Phi(r)$ the relative potential. The
cumulative mass contained within the sphere of radius $r$ is
\begin{equation}
\Ms(r)=\Ms\times\frac{s}{1+s},
\label{eq:Ms} 
\end{equation}
thus $\rs$ is the half-mass (spatial) radius.

The projected density at radius $R$ in the projection plane (see, e.g., Binney \& Tremaine 2008) is written as
\begin{equation}
\Sigs(R)=2\!\int_R^{\infty}{\rhos (r)r\over\sqrt{r^2-R^2}}dr
=\frac{\Ms}{\rs^2}\times\fstar(\eta),
\label{eq:sigs(R)}
\end{equation}
where $\eta\equiv R/\rs$, and $\fstar(\eta)$ is given in Appendix
\ref{App_A}. In particular,
\begin{equation}
\Sigs (R)\sim {\Ms\over\rs^2} \times\cases{
         \displaystyle{{1\over 4\eta}},
                                                     \qquad\,\,\, R\to 0,
         \cr\cr
         \displaystyle{{1\over 8\eta^3}},
                       \qquad R\to \infty.
         \cr\cr
         }
\label{eq:sigs(R)_asymp}
\end{equation}
Finally, the projected mass $\Mps(R)$ contained within the cylinder of radius $R$ is
\begin{equation}
\Mps(R)\equiv2\pi\!\int_0^R \Sigs (R')R'dR'=\Ms\!\times \gs(\eta),
\label{eq:proj_stellar_mass}
\end{equation}
where the function $\gs(\eta)$ is given in Appendix \ref{App_A}; as
expected, $\gs(\eta)\to 1$ for $\eta\to\infty$.  In particular,
  by setting $\gs=1/2$ we obtain the well-known result that
  $\reff\simeq 0.7447\,\rs$, where $\reff$ is the effective radius of
  the Jaffe profile.

\subsection{Total mass distribution}\label{sec:total_distr}

Following the considerations in the Introduction, the total (stars plus DM) mass density profile is 
\begin{equation}
\rhog(r)= {\MR\rhon\over s^2(\xi+s)}, \qquad \xi \equiv {\rg\over\rs},
\label{eq:rhog}
\end{equation}
where $\MR$ is a dimensionless factor, and $\rg$ is the galaxy scale
length. At variance with the stellar profile, the total mass is
divergent, so that now $\MR$ cannot be defined as the ratio of the
total-to-stellar mass (as in CZ18) but, more appropriately, as a {\it
  density} ratio. For example, from eqs. (\ref{eq:rhos_psis}) and
(\ref{eq:rhog}), it follows that $\MR=\xi\rhog(r)/\rhos(r)$ for $r\to
0$. In turn this means that the obvious request
$\rhog(r)\geq\rhos(r)$, when considered for $r\to 0$, forces
$\MR\geq\xi$. In Section \ref{sec:DM_distr} we shall fully address the
problem of the positivity of the DM density profile, obtained as the
difference $\rhoDM(r)=\rhog(r)-\rhos(r)$, over the whole radial range.

The cumulative mass distribution associated with the total galaxy density is
\begin{equation}
\Mg(r)=\Ms\MR\ln\frac{\xi+s}{\xi},
\label{eq:Mg}
\end{equation}
and diverges logarithmically. Incidentally, eq.~(\ref{eq:Mg}) allows
for different interpretations of the parameter $\MR$, in terms of
cumulative masses inside some prescribed radius; for example,
$\MR=\Mg(\rg)/(\Ms\ln 2)$.

From a simple integration, the projected galaxy total density profile
is
\begin{equation}
\Sigg(R)=\frac{\Ms}{\rs^2}\,\frac{\MR}{\xi^2}\times\fgal(\eta),
\label{eq:Sigg(R)}
\end{equation}
where now $\eta\equiv R/\rg$, and $\fgal(\eta)$ is given in Appendix
\ref{App_A}, and
\begin{equation}
\Sigg(R)\sim\frac{\Ms\MR}{\rs^2\,\xi^2}\times\cases{
         \displaystyle{\frac{1}{4\eta}},
                                                     \qquad\quad\, R\to 0,
         \cr\cr
         \displaystyle{\frac{1}{2\pi\eta^2}},
                       \qquad R\to \infty.
         \cr\cr
         }
\end{equation}
The projected mass within a cylinder of radius $R$ is 
\begin{equation}
\Mpg(R)=\Ms\MR\times\ggal(\eta), 
\label{eq:proj_total_mass}
\end{equation}
where the function $\ggal(\eta)$ is given in Appendix \ref{App_A}, and
$\ggal(\eta)\sim\ln\eta$ for $\eta\to\infty$. The gravitational
potential can be easily determined; even though the total mass is
infinite, yet the normalization value at infinity can still be assumed
equal to zero, as the density profile at large radii is steeper than
$r^{-2}$. The resulting relative potential associated with the density
$\rhog(r)$ is
\begin{equation}
\Psig(r)=\Psin\MR\times\left(\frac{1}{\xi}\ln\frac{\xi+s}{s}+\frac{1}{s}\ln\frac{\xi+s}{\xi}\right)\!.
\label{eq:Psig}
\end{equation}
Note that the first term is nothing else that the rescaled Jaffe
potential in \cref{eq:rhos_psis}, and dominates in the central
regions, while the second term becomes dominant at large radii.

As a BH of mass $\Mbh =\mu\Ms$ is added at the center of the galaxy, the total
mass profile is $\MT(r)=\Mg(r)+\Mbh$; then the circular speed is given
by
\begin{equation}
\vc^2(r)=\frac{\Psin}{s}\times\left(\MR\ln\frac{\xi+s}{\xi}+\mu\right)\!,
\end{equation}
and in the very external regions the circular velocity falls to zero
as
\begin{equation}
\vc^2(r)\sim\Psin\MR\,\,\frac{\ln s}{s}.
\end{equation}

\subsection{The dark matter distribution: positivity and monotonicity}\label{sec:DM_distr}

As already done for JJ models, we first study the conditions for the
positivity and radial monotonicity of the DM halo density profile
$\rhoDM(r)=\rhog(r)-\rhos(r)$. While the request of positivity for
$\rhoDM(r)$ is natural, we recall that monotonicity of the density is
necessary for the positivity of the phase-space DF (Ciotti \&
Pellegrini 1992, hereafter CP92; see also Sect. \ref{sect_DF}).

\begin{figure*}
 \centering
  \includegraphics[scale=0.835]{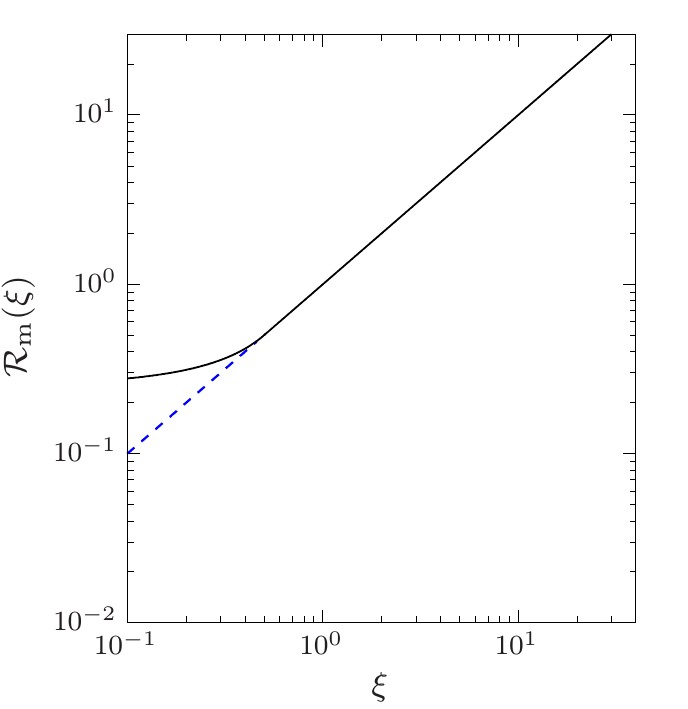}
  \includegraphics[scale=0.835]{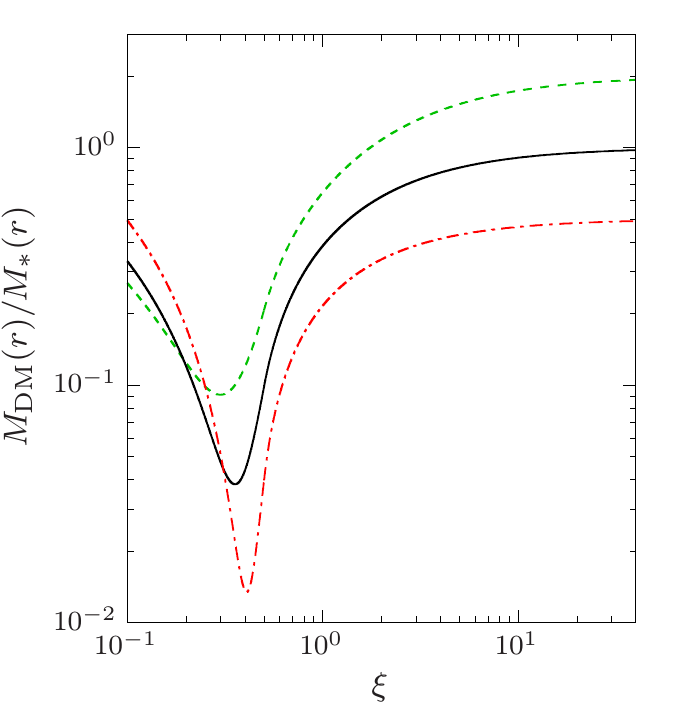}
  \includegraphics[scale=0.835]{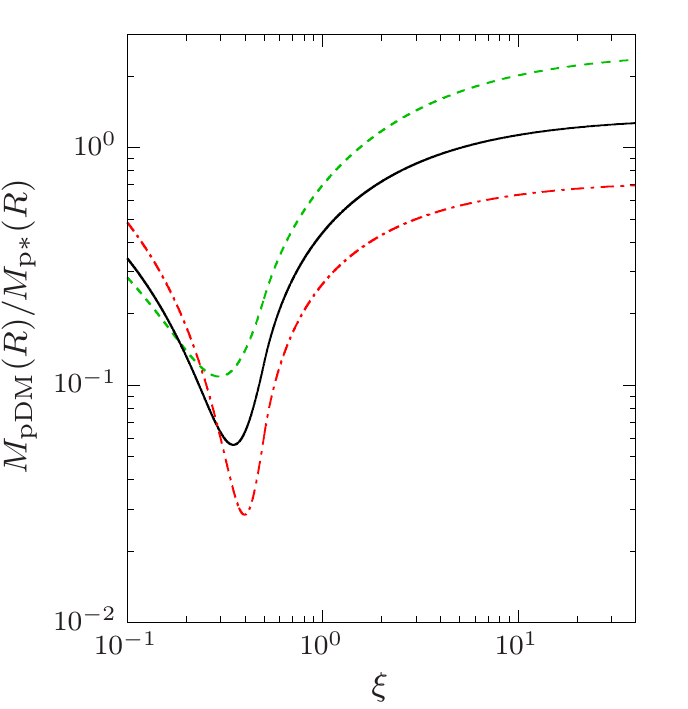}
 \vspace*{-1mm}
 \caption{Left panel: the function $\Rm(\xi)$, as given by
   \cref{eq:pos}. Only models in the region above the black solid line
   have a DM halo with a nowhere negative $\rhoDM(r)$. Central panel:
   the minimum value of the volumic DM-to-stellar mass ratio, given in
   \cref{eq:MD/Ms}, inside a sphere of radius $r=0.5\rs, \rs, 2\rs$
   (red dash-dotted, black solid and green dashed lines, respectively),
   as a function of $\xi$. Right panel: the minimum value of the
   projected DM-to-stellar mass ratio, given in \cref{eq:MpDM/Mps},
   inside the circle of radius $R=0.5\reff, \reff, 2\reff$ (red
   dash-dotted, black solid and green dashed lines, respectively), as a
   function of $\xi$. See Sect. \ref{sec:DM_distr} for more details.}
\label{fig:trittico}
\end{figure*}

Not all values of $\MR$ and $\xi$ are compatible with a nowhere negative
distribution 
\begin{equation}
\rhoDM(r)=\frac{\rhon}{s^2}\times\left[\,\frac{\MR}{\xi+s}-\frac{1}{{(1+s)^2}\,} \right]\!.
\label{eq:rhoDM}
\end{equation}
Note that in general the DM-to-stellar mass ratio $\rhoDM (r)/\rhos (r)$ depends on $r$.

In Appendix \ref{App_B} we determine the condition on $\MR$ and $\xi$
such that $\rhoDM(r) \ge 0$ for $r\ge 0$, obtaining
\begin{equation}
\MR\,\ge\,\Rm(\xi)\,=\,\cases{
            \displaystyle{\frac{1}{4(1-\xi)}}\,,
                                                     \qquad 0<\xi\leq {1 \over 2}\,,
         \cr\cr
         \displaystyle{\xi},
                       \qquad\qquad\qquad\;\, \xi\geq\frac{1}{2}\,.
         } 
\label{eq:pos}
\end{equation} 
A DM halo of a model with $\MR=\Rm(\xi)$ is called a {\it minimum
  halo}. Notice that the condition in \cref{eq:pos} for $\xi\geq 1/2$,
i.e. $\MR\geq\xi$ is coincident with that obtained in
Sect. \ref{sec:total_distr} from the preliminary analysis near the
center; instead, for $0<\xi<1/2$, the condition is more stringent (see
the dashed line in Fig. \ref{fig:trittico}, left panel). This means
that values of $(\xi, \MR)$ between the dashed and solid lines in
Fig. \ref{fig:trittico} correspond to $\rhoDM(r)$ that becomes
negative off-center. We stress that the formulae in the paper (if not
differently stated) apply to $\xi>0$, even if realistic cases (i.e., a
total density profile shallower than the stellar one) are obtained for
$\xi\geq 1$.

The positivity of $\rhoDM(r)$ is just a first condition for the
viability of the model. A second request, based on dynamical arguments
(see Section \ref{sec:NC_SC}), is the monotonicity of $\rhoDM$ as a
function of radius, and this reduces to the determination of the
minimum value $\Rmon$ so that $d\rhoDM(r) /dr \leq 0$. The explicit
discussion is deferred to Appendix \ref{App_B}, where we prove that,
for $\xi\geq 1/2$, positivity and monotonicity of $\rhoDM$ coincide
(in analogy with what found for JJ models).

We discuss now the relative trend of DM and stars, both at large radii
and near the center, as a function of $\MR$ and $\xi$.  For
$r\to\infty$ it is easy to show that
\begin{equation}
\rhoDM(r)\sim\frac{\rhon\MR}{s^3}, \qquad r\to\infty,
\label{eq:rhoDM_inf}
\end{equation}
and so the DM is dominant over the stars. Close to the center, instead,
\begin{equation}
\rhoDM(r)=\frac{\rhon}{\xi}\times\left(\frac{\MR-\xi}{s^2}+\frac{2\xi^2-\MR}{\xi s}+\frac{\MR-3\xi^3}{\xi^2}+.\,.\,.\,.\right)\!,
\label{eq:rhoDM_near_zero}
\end{equation}
so the trend depends on the values of $\MR$ and $\xi$. From
eqs. (\ref{eq:rhoDM_near_zero}) and (\ref{eq:pos}) it follows that, in
non-minimum halo models, $\rhoDM(r)\propto r^{-2}$, so the DM and
stellar densities are locally proportional. In the minimum halo models
we have $\rhoDM(r)\propto r^{-2}$ for $\xi<1/2$,
$\rhoDM(r)\propto{\rm constant}$ for $\xi=1/2$, and finally
$\rhoDM(r)\propto r^{-1}$ for $\xi>1/2$. In particular, in the latter
case, when $\MR=\Rm(\xi)=\xi$, one has
\begin{equation}
\rhoDM(r)\sim\rhon\,\frac{2\xi-1}{\xi s}, \quad r\to 0,
\label{eq:rhoDM_zero}
\end{equation} 
and so $\rhoDM(r)$ is centrally shallower than $\rhos(r)$. We shall
discuss an interesting application of eqs. (\ref{eq:rhoDM_inf}) and (\ref{eq:rhoDM_zero}) at the end of
this Section.

We evaluate now the relative amount of dark and visible mass within a
prescribed spatial radius. The minimum value for this quantity is
derived from eqs. (\ref{eq:Ms}) and (\ref{eq:Mg}) as
\begin{equation}
{\MD(r)\over\Ms(r)}\geq\Rm(\xi)\,\frac{1+s}{s}\ln\frac{\xi+s}{\xi}-1,
\label{eq:MD/Ms}
\end{equation}
where $\MD(r)=\Mg(r)-\Ms(r)$. In Fig. \ref{fig:trittico} (middle
panel) the mass ratios corresponding to three representative values of
$r$ are shown as a function of $\xi$. Note that, for fixed values of
$s$, the function at the r.h.s. of \cref{eq:MD/Ms} tends to $s$ for
$\xi\to\infty$. A similar behavior is obtained for the ratio of the
projected DM-to-visible mass within an aperture $R$; from
eqs. (\ref{eq:proj_stellar_mass}) and (\ref{eq:proj_total_mass}) one
has
\begin{equation}
\frac{\MpDM(R)}{\Mps(R)}\geq\Rm(\xi)\,\frac{\ggal(\eta/\xi)}{\gs(\eta)}-1,
\label{eq:MpDM/Mps}
\end{equation}
where $\eta=R/\rs$. In Fig. \ref{fig:trittico} (right panel), we plot
this quantity as a function of $\xi$ for three representative values
of the aperture radius: $\reff/2$, $\reff$, and $2\reff$. As for JJ
models, the resulting functions of $\xi$ are non monotonic.

We finally compare the DM halo profile in \cref{eq:rhoDM} with the NFW
profile (Navarro et al. 1997), which we rewrite as
\begin{equation}
\rhoNFW(r)=\frac{\rhon\MRN}{f(c)s(\csih +s)^2},\quad f(c)=\ln(1+c) -{c\over 1+c},
\label{eq:rhoNFW_f(c)}
\end{equation}
where for a chosen radius $\rt$, we define
$\MRN\equiv M_{\rm NFW}(\rt)/\Ms$, $\csih \equiv r_{\rm NFW}/\rs$ is
the NFW scale length in units of $\rs$, and $c\equiv\rt/r_{\rm
  NFW}$. By construction, $\rhoDM(r)$ and $\rhoNFW(r)$ at large radii
have the same behaviour. Moreover, at small radii, where
$\rhoNFW(r)\propto r^{-1}$, in the minimum halo case with $\xi>1/2$,
one also has that $\rhoDM(r)\propto r^{-1}$.
Remarkably, from eqs. (\ref{eq:rhoDM_inf}) and (\ref{eq:rhoDM_zero}) it follows that $\rhoDM(r)$ and
$\rhoNFW(r)$ can be made asymptotically {\it identical} in the outer
regions and near the center, just by imposing
\begin{equation}
\frac{\MRN}{f(c)}=\xi,\qquad\csih=\frac{\xi}{\sqrt{2\xi-1}}.
\label{eq:csiNFW}
\end{equation} 
Therefore, once a specific minimum halo galaxy model with
$\xi\geq 1/2$ is considered, and then $\rhon$ and $\xi$ are chosen,
eqs. (\ref{eq:rhoNFW_f(c)}) and (\ref{eq:csiNFW}) allow to determine
the NFW profile that best reproduces $\rhoDM(r)$ by tuning the value
of the ratio $\MRN/f(c)$. As pointed out in the Introduction, the
possibility to have a DM distribution very similar to the NFW profile
both at the center {\it and} in the outer regions makes J3 models an
improvement over JJ models, whose DM profile is necessarily more and
more discrepant from the NFW profile with increasing radii.

\begin{figure*}
 \centering
  \includegraphics[scale=1.02]{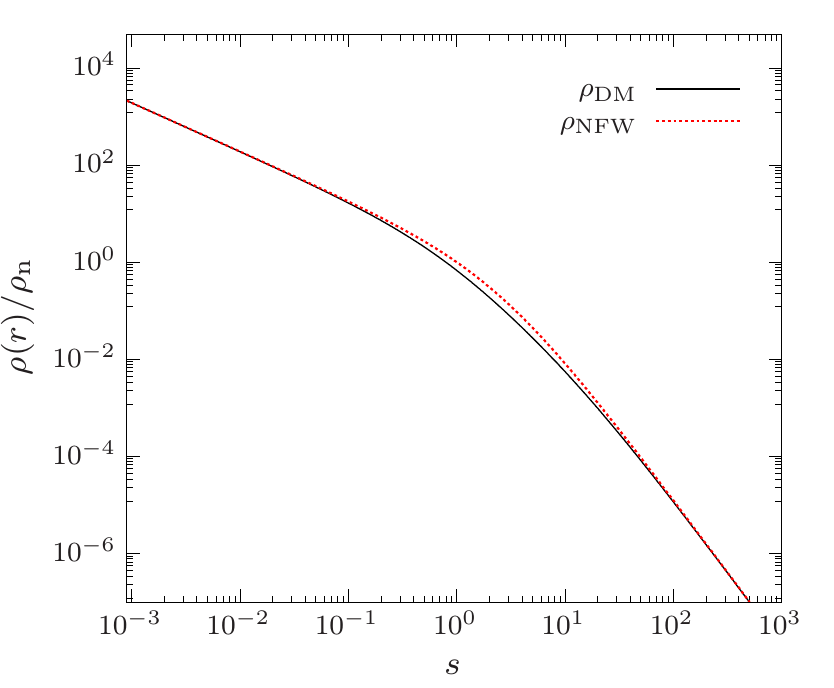}\quad\,
  \includegraphics[scale=1.02]{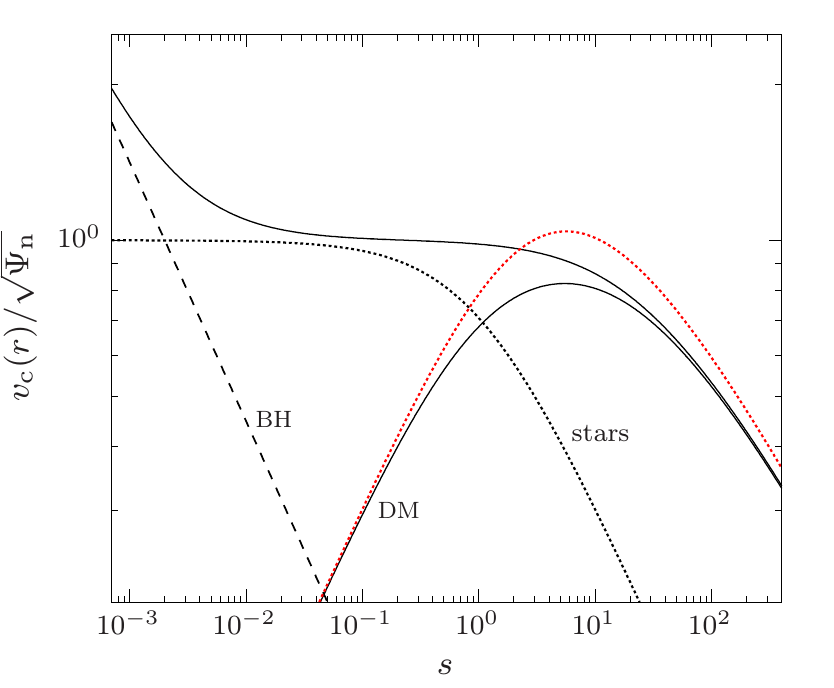}
 \vspace*{-4mm}
 \caption{Left panel: comparison between the minimum halo DM profile
   of J3 models (black solid line) and the NFW profile (red dotted
   line), for $\xi=13$ and $c=10$. Right panel: circular velocity
   profile for the same minimum halo galaxy model with $\mu=0.002$
   (solid line), with the separate contributions of BH, stars and the
   DM halo for the J3 model; the circular velocity profile associated
   with the density $\rhoNFW(r)$ in the left panel (red dotted line)
   is also shown. See Sect. \ref{sec:DM_distr} for more details.}
\label{fig:NFW_vc}
\end{figure*}

Figure \ref{fig:NFW_vc} (left panel) shows an example of how well a
NFW profile can reproduce a minimum halo $\rhoDM(r)$, when both
profiles are chosen in a cosmologically motivated way. In this figure
$\xi=13$ and $c=10$, that give, from \cref{eq:csiNFW},
$\MRN\simeq 20$, as expected on the scale of massive galaxies from
cosmological simulations and galaxy-halo abundance matching technique
(e.g., Bullock \& Boylan-Kolchin 2017). For this model
$r_{\rm NFW}=2.6\,\rs$ from eq.~(23), and the right panel in
Fig. \ref{fig:NFW_vc} shows the different contributions of the various
mass component to the circular velocity $\vc(r)$.

\section{The phase-space  distribution function}\label{sect_DF}

We now proceed to recover the OM phace-space DF for J3 models in
presence of a central BH.  Indeed, for OM models it is possible to
obtain lower bounds for the anisotropy radius without actually
recovering the DF (that only exceptionally can be expressed in terms
of elementary or known functions; see, e.g., CP92, Ciotti 1996, Ciotti
1999). We shall use, as for JJ models, a result presented in
CP92\footnote{The conditions in CP92 are a special case of a more
  general family of inequalities (see, e.g., de Bruijne et al. 1996,
  An \& Evans 2006, Ciotti \& Morganti 2009, 2010ab, van Hese et
  al. 2011, and references therein).}.

We therefore consider, for the stellar component, a DF of the family
\begin{equation}
f=f(Q),\quad Q\equiv \Er - {J^2\over 2\ra^2},
\label{eq:OM_param}
\end{equation}
(e.g., Binney \& Tremaine 2008), where $\ra$ is the anisotropy radius, 
and $\Er =\PsiT-v^2/2$ and $J$
are the relative energy and angular momentum modulus of each star (per
unit mass); moreover, $f(Q)=0$ for $Q<0$.  The total (relative)
gravitational potential is $\PsiT(r)=\Psig(r)+G\Mbh/r$, and from
\cref{eq:Psig} one has
\begin{equation}
\frac{\PsiT(r)}{\Psin}\equiv\psi(r) = 
\MR\left(\frac{1}{\xi}\ln\frac{\xi+s}{s}+\frac{1}{s}\ln\frac{\xi+s}{\xi}\right)\!+\frac{\mu}{s}.
\label{eq:PsiT}
\end{equation}
The anisotropy profile is given by
\begin{equation}
\beta(r)\equiv{1-{\stan^2(r)\over{2\srad^2(r)}}}={r^2\over{r^2+\ra^2}},
\label{eq:beta}
\end{equation}
where $\srad(r)$ and $\stan(r)$ are the radial and tangential components 
of the velocity dispersion tensor, respectively. The fully isotropic case is obtained 
for $\ra\to\infty$, while for $\ra=0$ the galaxy is supported by radial orbits only. For
finite values of $\ra$, instead, the velocity dispersion tensor becomes
isotropic for $r\to 0$, and fully radially anisotropic for $r\to\infty$. 

The phase-space DF of the stellar component reads
\begin{equation}
f(Q)=\frac{1}{\sqrt{8}\pi^2}\int_0^{Q}\!{d^2\!\varrho\over d\PsiT^2}\frac{d\PsiT}{\sqrt{Q-\PsiT}},
\label{eq:f(Q)}
\end{equation}
where
\begin{equation}
\varrho(r)\equiv\rhos(r)\times\left(1+{r^2\over\ra^2}\right)
\label{eq:varrho}
\end{equation}
is intended expressed in terms of $\PsiT$ by elimination of radius.

\subsection{Necessary and sufficient conditions for consistency}\label{sec:NC_SC}

Before the numerical reconstruction of the DF and the determination of
the critical value of $\ra$ for consistency, it is instructive to
study preliminarly the limitations on the anisotropy radius obtained
by the request of $f(Q) \geq 0$ over the accessible phase-space. The
CP92 {\it necessary condition} for the positivity of the DF of each
mass component of J3 models in the total potential is that
\begin{equation}
{d\varrho (r)\over dr}\leq 0,\quad{\rm[NC]}.
\label{eq:NC}
\end{equation}
Notice how this condition is actually independent of the presence of
other density components. Moreover, a {\it weak sufficient condition} for
consistency reads
\begin{equation}
{d\over dr}\!\left[{d\varrho(r)\over dr}{r^2\over\MT (r)}\right]\geq 0, \quad{\rm [WSC]},
\label{eq:WSC}
\end{equation}
where it is apparent that, at variance with the NC, the WSC depends
also on the radial density profile of the other
components. Summarizing, a model failing \cref{eq:NC} is certainly
inconsistent, while a model obeying \cref{eq:WSC} is certainly
consistent. Hence, the limitations obtained from the applications of
the two previous inequalities are expected to ``bracket'' the true
limitations on the model parameters, which can be only determined by
direct inspection of the DF.

Before embarking on the analysis of the consistency, we recall a
couple of points. The first concerns the effect of the central BH on
consistency. From \cref{eq:WSC} it follows that if 1) the investigated
density component satisfies the WSC for $\Mbh =0$, and 2) $d (r^2
d\varrho/dr)/dr\geq 0$, then the WSC is satisfied\footnote{The
  conditions 1) and 2) are special cases of a more general result of
  easy proof, i.e., the fact that \cref{eq:WSC} is necessarily true
  once it is separately true for $\varrho(r)$ in each mass component
  producing $\MT(r)$.} for arbitrary values of $\Mbh$. The second
consideration is about the effect of orbital anisotropy. In fact, the
investigation of the NC and WSC, and the study of the DF positivity in
\cref{eq:f(Q)}, all lead to consider inequalities that can be written
as
\begin{equation}
F+{G\over\sa^2}\geq 0, \quad \sa\equiv{\ra\over\rs},
\label{eq:FG_ineq}
\end{equation}
and that must hold over the domain $\cond$ spanned by the arguments of
the functions $F$ and $G$ For example, in case of the DF these two
functions are given by eq. (\ref{eq:f(q)}), with argument $q$; or, in
case of NC and WSC, by the radial functions obtained from
\cref{eq:varrho}. In the following we discuss the DF case. Equation
(\ref{eq:FG_ineq}) shows that all OM models can be divided in two
families. In the first case, when $F$ is nowhere negative over $\cond$
(e.g., in the case of a consistent isotropic DF), consistency is
obtained for
\begin{equation}
\sa \geq \sam \equiv \sqrt{{\rm max}\!\left[0,\,{\rm sup}_{\cond}\!\left(-{G\over F}\right)\right]}.
\label{eq:sa_minus}
\end{equation}
If $G \geq 0$, then $\sam =0$ and the system can be supported by
radial orbits only. In the second case, $F \geq 0$ over some subset
$\condp$ of $\cond$, and negative (or zero) over the complementary
subset $\condm$.  If also $G <0$ somewhere on $\condm$, then the model
is inconsistent.  If $G \ge 0$ on $\condm$ one must consider the lower
limit $\sam$ evaluated over $\condp$ as above, {\it and} the upper
limit
\begin{equation}
\sa \le \sap=\sqrt{{\rm inf}_{\condm}\!\left(-{G\over F}\right)},
\label{eq:sa_plus}
\end{equation}
over $\condm$: consistency is possible only if $\sam<\sap$.

The first application of the NC and WSC to J3 models concerns the
consistency of the DM halo.  For simplicity, we restrict to the
isotropic case, when \cref{eq:NC} shows the equivalence of the NC with
the request of monotonicity of $\rhoDM(r)$: remarkably, according to
the results in eqs. (\ref{eq:mon})-(\ref{eq:Mmon}), for $\xi\geq 1/2$
the NC is satisfied once just positivity is assured, i.e. for
$\MR\geq\xi$. The WSC for the isotropic DM halo with a central BH
requires in general a numerical study: following the considerations
after eq.~(30), in Appendix \ref{App_B} we prove that for $\xi\geq 1$
the WSC is satisfied once $\MR\geq\xi$. This means that for isotropic
DM halos of J3 models with a central BH and $\xi\geq 1$, the
requirements of positivity, monotonicity, and WSC for phase-space
consistency coincide, and are all satisfied once $\MR\geq\xi$.

We now apply the NC and WSC to the Jaffe stellar component of J3
models. We recall that the NC of a pure Jaffe model just reduces to
$\sa\geq 0$, while the WSC gives $\sa\geq\sam\simeq 0.1068$ (e.g.,
Ciotti 1999, CZ18). In Appendix \ref{App_B} we show that the WSC
always produces the case described by \cref{eq:sa_minus}, i.e. only
$\sam$ exists. When restricting to the case $\mu=0$ the limit on
anisotropy is independent of $\MR$, and by numerical solution of
\cref{eq:wsc_stellar} we obtain $\sam=\sam(\xi)$ (dotted red line in
Fig. \ref{fig:wsc}). At the opposite limit we have the BH dominated
case (see Appendix A of CZ18), with $\sam\simeq 0.31$, obviously
independent of $\xi$ and $\mu$. Again following the discussion after
\cref{eq:WSC}, from comparison with Fig. \ref{fig:wsc} we conclude that the stellar
component of J3 models is certainly consistent for $\sa \geq 0.31$.

Summarizing, for J3 models this preliminary analysis reveals that the
presence of a diffuse halo allows for the possibility of more radial
orbits, while a concentrated halo requires a more isotropic velocity
dispersion tensor for stellar consistency (for similar results see,
e.g., Ciotti 1996, 1999; CMZ09; CZ18).

\begin{figure}
 \centering
  \includegraphics[scale=1.11]{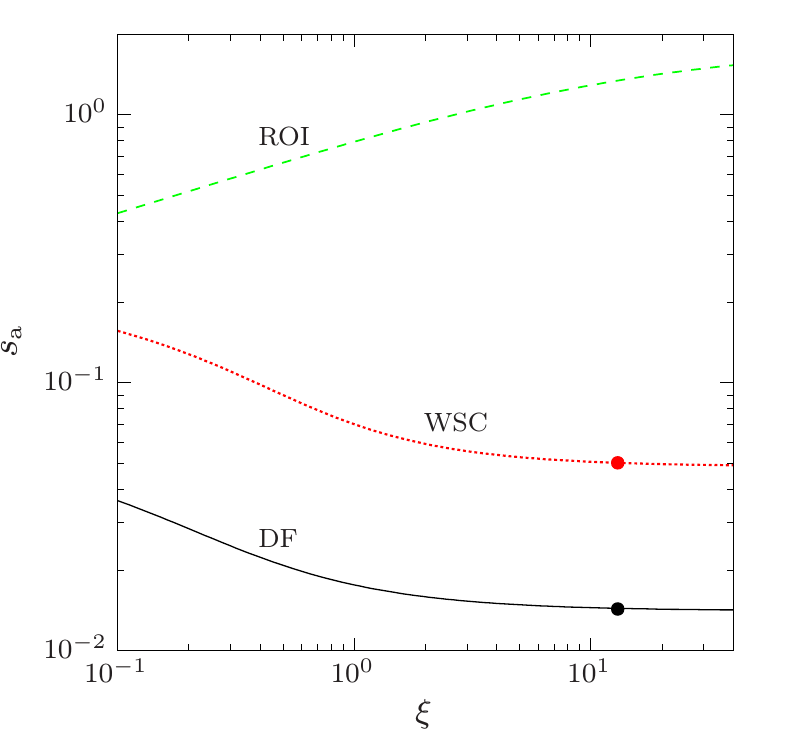}
 \vspace*{-5mm}
 \caption{ Different limitations on the anisotropy radius
   $\sa=\ra/\rs$ of the stellar component of J3 models, as a function
   of $\xi=\rg/\rs$. The lines refer to $\mu=0$, i.e. in absence of
   the central BH. The black solid line and the red dotted line
   represent the minimum value of $\sa$ obtained directly from the DF
   and from the WSC (see eq. [\ref{eq:wsc_stellar}]), respectively,
   while the green dashed curve represents the fiducial lower limit of
   $\sa$ to prevent the onset of Radial Orbit Instability (see
   Sect. \ref{sec:stability}). The circles correspond to models with
   $\xi=13$, for which we find $\sam\simeq 0.0143$ (DF) and
   $\sam\simeq 0.0502$ (WSC). For reference, the minimum value for consistency
   for the Jaffe model is $\sam \simeq 0.02205$.}
\label{fig:wsc}
\end{figure}

\subsection{Explicit phase-space DF}\label{sec:explicit_DF}

With the introduction of the dimensionless potential
$\psi=\PsiT/\Psin$, and the normalized augmented density
$\tilde\varrho=\varrho/\rhon$, \cref{eq:f(Q)} becomes
\begin{eqnarray}
f(q)&=&{\rhon\over\sqrt{8}\pi^2\Psin^{3/2}}
     \int_0^{q}{d^2\tilde\varrho\over d\psi^2}{d \psi \over\sqrt{q-\psi}}\cr\cr\cr
    &=&{\rhon\over\sqrt{8}\pi^2\Psin^{3/2}}
       \left[U(q)+{V(q)\over\sa^2}\right]\!,
\label{eq:f(q)}
\end{eqnarray}
where $q\equiv Q/\Psin$, and, from \cref{eq:PsiT},
$0\leq q\leq\infty$. Note how the resulting expression belongs to the
family in \cref{eq:FG_ineq}.

In the case of a pure stellar Jaffe model the functions $U(q)$ and
$V(q)$ can be obtained analytically, as also for a Jaffe model with a
central dominant BH; moreover, the function $\tilde\varrho (\psi)$ for
JJ models with a central BH can be obtained analitically (e.g., see
Appendix C of CZ18, and references therein). Unfortunately, for J3
models $U(q)$ and $V(q)$ cannot be obtained in terms of elementary
functions. Therefore, in the following discussion we shall proceed
with the numerical integration of eq. (\ref{eq:f(q)}) for a selected
choice of the model parameters, changing the integration variable from
potential to radius, so that the integrand can be written as an
explicit analytical function.

First, we determine numerically the lower limit on $\sa$ for
consistency by inspection of the functions $U(q)$ and $V(q)$. Note
that in absence of the central BH ($\mu=0$), from \cref{eq:PsiT} the
variable $q$ can be further scaled as $\qt=q/\MR$, and the quantity
$\MR^{-3/2}$ can be explicitily factored out in the functions $U(q)$
and $V(q)$ (see also eq. [C4] in CZ18). In particular, for models
without a central BH, the position of the maximum in
\cref{eq:sa_minus} depends on $\qt$, and the value of $\sam$ is
independent of $\MR$. It is numerically found that $U(q)\geq 0$, so
that \cref{eq:sa_minus} applies and only $\sam$ exists: the black
solid line in Fig. \ref{fig:wsc} shows the corresponding
$\sam(\xi)$. At fixed $\xi$, anisotropy values $\sa\geq\sam$
correspond to a positive DF. Notice how the shape of the critical
consistency curve parallels the WSC condition (red dotted line). For
reference, the black circle at $\xi=13$ marks the minimum anisotropy
radius ($\sam\simeq 0.0143$) for our representative J3 model. From
Fig. \ref{fig:wsc} it is also apparent how the effect of a
concentrated DM halo reduces the ability of the stellar component to
sustain radial orbits, a common property of the OM models, confirming
the trend obtained from the WSC. Here we mention a point of little
practical interest, but quite relevant conceptually. Indeed, in CZ18
was shown that for the single component Jaffe model the OM DF
requires, for consistency, $\sam \simeq 0.02205$, and so one could
argue that the purely radial Jaffe model does not exist. However, the
analytical DF for this particular model is positive, thus showing that
the purely radial case is a singular limit for the OM DF: indeed,
eqs. (C8) and (C10) in CZ18 show that the Jaffe stellar component of
J3 models can be supported by radial orbits only. In any case this
situation has not practical interest, as illustrated in
Sect. \ref{sec:stability}.

\begin{figure}
 \centering
  \includegraphics[scale=1.11]{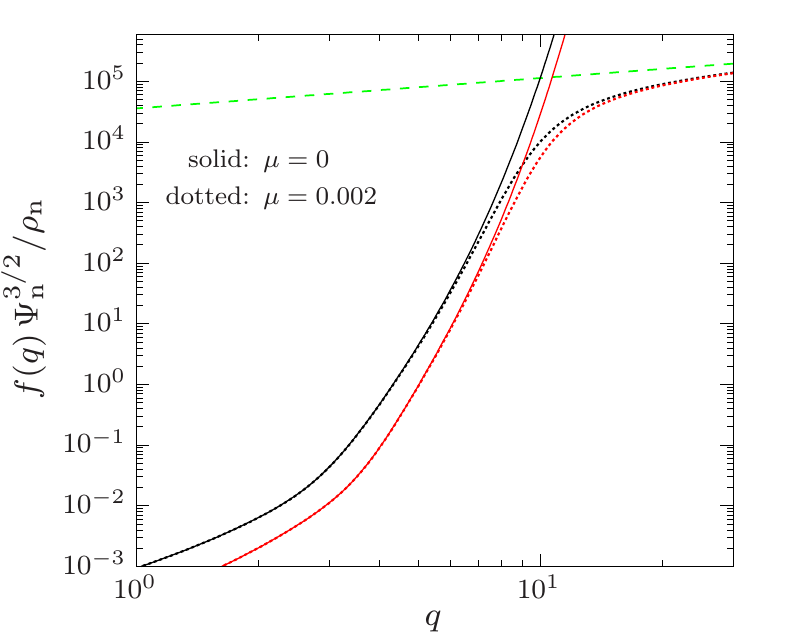}
  \includegraphics[scale=1.11]{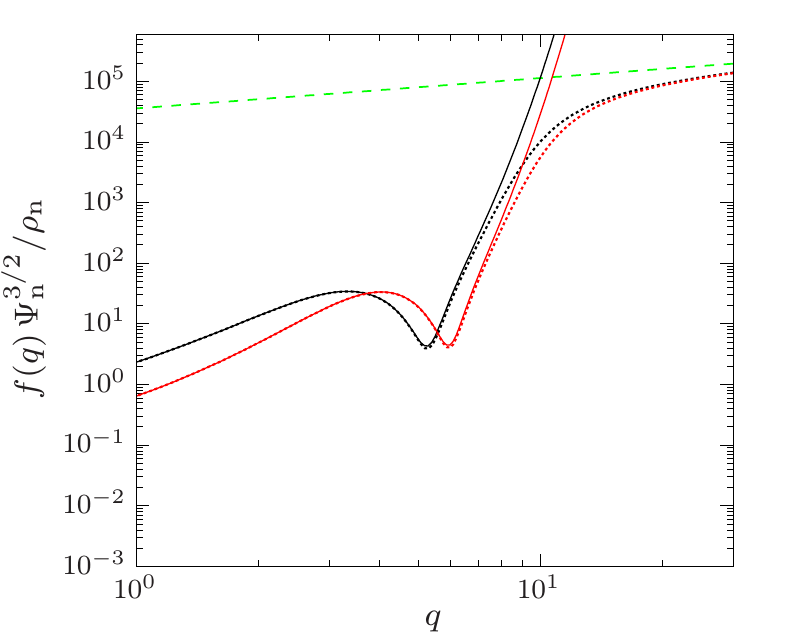}
 \vspace*{-5mm}
 \caption[hello]{The DF of the stellar component for a minimum halo
   galaxy model with $\MR=\xi=10$ (black lines) and $\MR=\xi=20$ (red
   lines), in the isotropic (top panel) and anisotropic (bottom panel,
   $\sa=0.02$) cases. The DFs are shown with and without (dotted and
   solid lines, respectively) the effect of the central BH (with
   $\mu=0.002$). The green dashed line shows the DF of the Jaffe model
   in the BH dominated case (see Sect. \ref{sec:explicit_DF}).}
\label{fig:DF}
\end{figure}

Figure \ref{fig:DF} shows the numerically recovered DF of the stellar
component of a selection of J3 models, namely the minimum halo models
with $\MR=\xi=10$ (black lines) and $\MR=\xi=20$ (red lines), in the
isotropic (top panel) and anisotropic (bottom panel, $\sa=0.02$)
cases; the DFs are shown with and without the effect of the central
BH. For illustration, also the BH dominated DF (green dashed line) is
shown. It is clear how at high relative energies the DF of the J3
models with a central BH is matched by the BH dominated DF, and how
the values of the isotropic and anisotropic DFs become coincident. As
in JJ models, the DFs with the central BH are lower at high relative
energies than in the analogous models without the central BH, the same
happens at low relative energies for models with heavier and extended
halos. These can be qualitatively interpreted when considering that
the DF values are expected to be inversely proportional to the cube of
velocity dispersion, so that high velocity dispersions are expected to
correspond to low values of the DF (cfr. with
Fig. \ref{fig:sigma_3D_p}).  We also notice that the curves relative
to DFs in the strongly anisotropic cases behave (qualitatively) as the
DFs of other OM models discussed in Ciotti \& Lanzoni 1997 (Fig. 2),
Ciotti 1999 (Figs. 2 and 3), CMZ09 (Fig. 3), and CZ18 (Fig. 3). In
practice, in OM models small values of $\sa$ lead to a depression of
the DF at intermediate energies, where model inconsistency finally
sets in when $\sa$ drops below the consistency limit.

\section{Jeans equations with OM anisotropy}\label{sect_JEs}

The Jeans equations for spherical systems with general anisotropy has
been discussed in Binney \& Mamon (1982), and in the OM models the
formal solution can be written as
\begin{eqnarray}
\rhos (r)\srad^2(r)&=&{G\over r^2 +\ra^2}
             \int_r^{\infty}\!\rhos(x)\MT(x)\!\left(1+{\ra^2\over x^2}\right)\!dx\cr\cr\cr
            &=&\rhon\Psin\,{{\cal{A}}(s)+ \sa^2\,{\cal{I}}(s)\over s^2+\sa^2},
\label{eq:rhosigma2}
\end{eqnarray}
where in our case
\begin{equation}
 {\cal{I}}(s) =\MR\,\It (s) +\mu\,\Ibh (s),
\end{equation}
\begin{equation}
 {\cal{A}}(s) =\MR\,\At(s)+\mu\,\Abh (s).
\end{equation}
In the two radial functions above the parameters $\MR$ and $\mu$ have
been explicitely factorized; for $\sa\to\infty$ the solution of the
fully isotropic case is recovered, while for $\sa=0$ one reduces to
the purely radial case.

\subsection{The velocity dispersion profile}

The BH contribution to the velocity dispersion profile is given by
\begin{equation}
\Ibh(s) = {12s^3+6s^2-2s+1\over 3s^3(1+s)}+4\ln\frac{s}{1+s},
\end{equation}
\begin{equation}
\Abh(s)={1+2s\over s(1+s)}+2\ln\frac{s}{1+s}.
\end{equation}
The galaxy contribution is obtained with an integration by parts, and
it can be written as
\begin{equation}
\It(s)\,=\,\Ibh(s)\ln\frac{\xi+s}{\xi}+\,\Fiso(\xi,s),
\end{equation}
\begin{equation}
\At(s)\,=\,\Abh(s)\ln\frac{\xi+s}{\xi}+\,\Frad(\xi,s),
\end{equation}
in which
\begin{equation}
\Fiso(\xi,s)\equiv\cases{
         \displaystyle{\frac{9\xi^2+3\xi+1}{3\xi^3}\ln\frac{\xi+s}{s}+\frac{1}{\xi-1}\ln\frac{\xi+s}{1+s}\,+}
         \cr\cr
         \displaystyle{-\frac{2(1+3\xi)s-\xi}{6\xi^2s^2}-4\Hcsi(\xi,s)},
         \cr\cr\cr
         \displaystyle{\frac{13}{3}\ln\frac{1+s}{s}-\frac{2s^2+7s-1}{6s^2(1+s)}-4\Hcsi(1,s)},
         }
\label{eq:Fiso}
\end{equation}
\begin{equation}
\Frad(\xi,s)\equiv\cases{
            \displaystyle{\frac{1}{\xi}\ln\frac{\xi+s}{s}+\frac{1}{\xi-1}\ln\frac{\xi+s}{1+s}-2\Hcsi(\xi,s)},
         \cr\cr\cr
         \displaystyle{\ln\frac{1+s}{s}+\frac{1}{1+s}-2\Hcsi(1,s)},
         }
\label{eq:Frad}
\end{equation}
where in eqs. (\ref{eq:Fiso}) and (\ref{eq:Frad}) the first formula
holds for $\xi\ne 1$, and the second for $\xi=1$; the function
$\Hcsi(\xi,s)$ is given in Appendix \ref{App_C}.

An insight into the behavior of $\srad(r)$ is given by the expansion
for $r\to \infty$ and $r\to 0$ of the obtained formulae. We begin with
the outer galaxy regions, where
\begin{equation}
\Ibh(s)=\frac{1}{5s^5}+\mathcal{O}\!\left(\frac{1}{s^6}\right)\!, \quad \Abh(s)=\frac{1}{3s^3}+\mathcal{O}\!\left(\frac{1}{s^4}\right)\!,
\end{equation}
and, from expansion of \cref{eq:Hcsi}, we obtain
\begin{equation}
\It(s)=\frac{\ln s}{5s^5}+\frac{1-5\ln\xi}{25s^5}+\mathcal{O}\!\left(\frac{\ln s}{s^6}\right)\!, 
\label{eq:It(s)}
\end{equation}
\begin{equation}
\At(s)=\frac{\ln s}{3s^3}+\frac{1-3\ln\xi}{9s^3}+\mathcal{O}\!\left(\frac{\ln s}{s^4}\right)\!.
\label{eq:At(s)}
\end{equation}

From eqs. (\ref{eq:rhosigma2}), (\ref{eq:It(s)}) and (\ref{eq:At(s)}), it follows that at large radii
\begin{equation}
\srad^2(r)\sim\frac{\Psin\MR\ln s}{s}\times\cases{
            \displaystyle{\frac{1}{5}},
                       \quad\;\;\; \sa=\infty,
         \cr\cr
         \displaystyle{\frac{1}{3}},
                                                     \quad\quad\! \sa<\infty.
         }
\end{equation}
Therefore, at the lowest order, $\srad(r)$ in the outer regions is
independent of $\xi$, and the asymptotic formula is similar to that of
JJ models (which just differs in the absence of the function $\ln s$).

The other important region for observational and theoretical works is
the galaxy center: here the velocity dispersion profile is dominated
by the BH contribution, with
\begin{equation}
\Ibh(s)=\frac{1}{3s^3}+\mathcal{O}\!\left(\frac{1}{s^2}\right), \quad \Abh(s)=\frac{1}{s}+\mathcal{O}(\ln s),
\label{eq:Ibh_Abh}
\end{equation}
while
\begin{equation}
\It(s)=\frac{1}{2\xi s^2}+\mathcal{O}\!\left(\frac{1}{s}\right)\!, \quad \At(s)=-\,\frac{\ln s}{\xi}+\mathcal{O}(1).
\label{eq:It_At_zero}
\end{equation}
Hence, as $r\to 0$, 
\begin{equation}
\srad^2(r)\sim\frac{\Psin\mu}{s}\times\cases{
            \displaystyle{\frac{1}{3}},
                       \qquad \sa>0,
         \cr\cr
         \displaystyle{1},
                                                     \quad\quad\; \sa=0.
         }
\label{eq:BH_dom}
\end{equation}
If the central BH is absent, instead, one has 
\begin{equation}
\srad^2(r) \sim \frac{\Psin\MR}{\xi}\times\cases{
            \displaystyle{\frac{1}{2}},
                       \qquad\quad\;\;\; \sa>0,
         \cr\cr
         \displaystyle{-\ln s},
                                                     \qquad\, \sa=0,
         }
\label{eq:rho_srad_noBH}
\end{equation}
and so, with the exception of the purely radial case,
\begin{equation}
\srad^2(0)=\frac{\Psin\MR}{2\xi}.
\label{eq:srad2_0}
\end{equation}
We notice that as a check of the obtained asymptotic formulae, we also
expanded the integral in eq.~(\ref{eq:rhosigma2}) in the relevant
regimes, recovering eqs. (\ref{eq:It(s)})-(\ref{eq:At(s)}) and
eqs. (\ref{eq:Ibh_Abh})-(\ref{eq:It_At_zero}).

\begin{figure}
 \centering
  \includegraphics[scale=1.11]{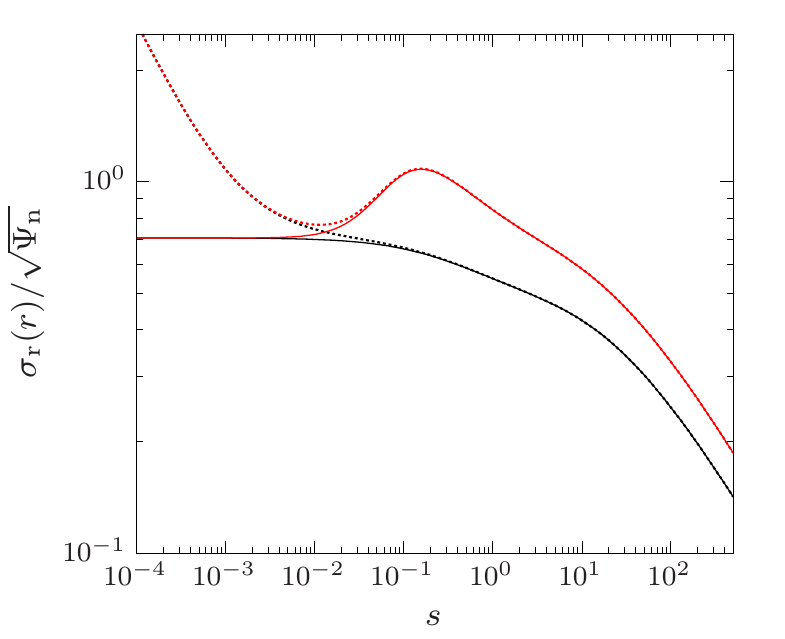}
  \includegraphics[scale=1.11]{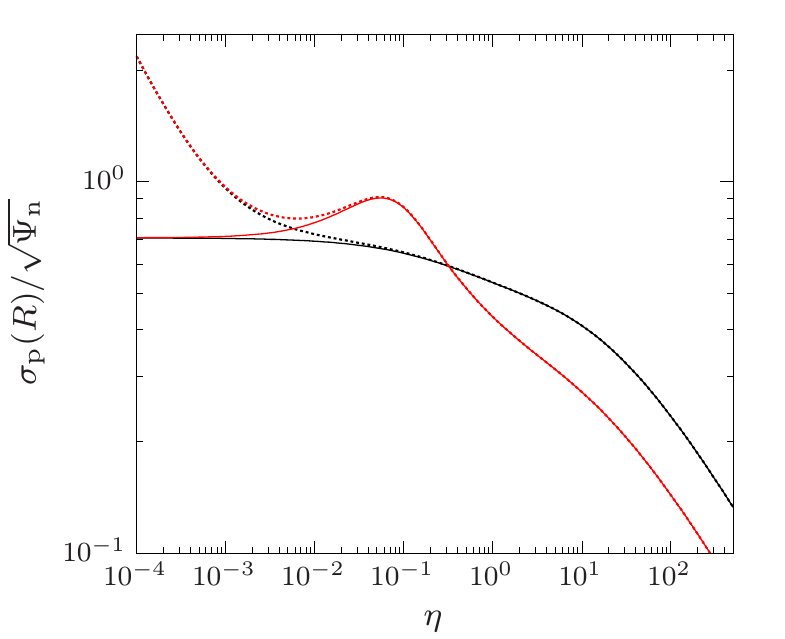}
 \vspace*{-5mm}
 \caption{Top panel: radial trend of $\srad$ of the stellar component
   for a minimum halo model with $\MR=\xi=13$. Black lines refer to
   the isotropic case, red lines show the quite anisotropic case with
   $\sa=0.1$. Bottom panel: radial trend of the projected stellar
   velocity dispersion $\sigp$ versus $\eta=R/\rs$ for the same model
   in the top panel. In both panels, the radial trends are shown with
   and without (dotted and solid lines, respectively) the effect of
   the central BH (with $\mu=0.002$).}
\label{fig:sigma_3D_p}
\end{figure}

The properties of $\srad(r)$ are illustrated in
Fig. \ref{fig:sigma_3D_p} (top panel) for the representative J3 models
with $\MR=\xi=13$. In particular, the effects of the central BH, of
the DM halo, and of orbital anisotropy can be clearly seen at large
radii, where the radially anisotropic $\srad(r)$ (red lines) are above
those in the corresponding isotropic cases, a well known consequence
of the OM parametrization. Note also how the isotropic and anisotropic
profiles coincide in the central regions, in accordance with the
analytical results in eqs. (\ref{eq:BH_dom}) and
(\ref{eq:rho_srad_noBH}).

\subsection{Projected velocity dispersion}

The projected velocity dispersion profile associated with a general
anisotropy function $\beta(r)$ is given by
\begin{equation}
\Sigma_*(R)\sigp^2(R)= 2\int_R^{\infty}\!{\left[{1-\beta(r)}
{R^2\over{r^2}}\right]}{{\rhos(r)\srad^2(r)rdr}\over{\sqrt{r^2-R^2}}},
\label{eq:rhosigma2p}
\end{equation}
(e.g., Binney \& Tremaine 2008), where, in the case of OM anisotropy, $\beta (r)$ is defined in \cref{eq:beta}.

Unsurprisingly the projection integral cannot be evaluated
analytically for J3 models in terms of elementary functions. However,
using eq. (\ref{eq:rhosigma2}), and changing the order of integration,
we find that, for a general OM model, \cref{eq:rhosigma2p} can be rewritten as
\begin{equation}
\Sigma_*(R)\sigp^2(R)=G\!\int_R^{\infty}\!\K(r,R)\rhos(r)\MT(r)\!\left(1+\frac{\ra^2}{r^2}\right)\!dr,
\label{eq:kernel_K}
\end{equation} 
where
\begin{equation}
\K(r,R)=\frac{2\ra^2+R^2}{(\ra^2+R^2)^{3/2}}\,\atan\,\sqrt{\frac{r^2-R^2}{\ra^2+R^2}}-\frac{R^2\sqrt{r^2-R^2}}{(\ra^2+R^2)(\ra^2+r^2)}.
\label{K_explicit}
\end{equation}
The special case $\ra=0$ {\it and} $R=0$ can be treated directly in
\cref{eq:rhosigma2p}. Equations
(\ref{eq:kernel_K})-(\ref{K_explicit}), although seem rather
complicated, actually reduce the dimensionality of the integral
(\ref{eq:rhosigma2p}) from two to one. This is an useful property in
numerical works, avoiding the task of the computation of the
two-dimensional integral (\ref{eq:rhosigma2p}). All the relevant
properties of $\sigp(R)$ are illustrated in the bottom panel of
Fig. \ref{fig:sigma_3D_p}. As expected, in the outer regions the
radially anisotropic profiles are below those in the corresponding
isotropic cases, a natural result due to the projection effect on the
radial orbit population. More quantitatively, the behaviour of
$\sigp(R)$ at large radii can be described by considering the leading
term of the asymptotic expansion of the integral in
\cref{eq:rhosigma2p} for $R\to\infty$. There are no mathematical
difficulties; the only care required is to take into account the
effect of the radial anisotropy, and to distinguish two different
cases, i.e., the isotropic case and any other model with finite
$\sa$. We obtain:
\begin{equation}
\sigp^2(R)\sim\frac{8\Psin\MR\ln\eta}{15\pi\eta}\times\cases{
            \displaystyle{1},
                       \qquad\;\, \sa=\infty,
         \cr\cr
         \displaystyle{\frac{1}{3}},
                                                     \qquad\, \sa<\infty,
         }
\end{equation}
where $\eta\equiv R/\rs$. Notice the similarity of this result with
that obtained for JJ models (see eq. [53] of CZ18). The numerical
coefficients are identical, but now there is an additional logarithmic
factor, and the BH mass ($\mu$) does not appear because the total mass
is infinite. Of course, also the coefficient $\MR$ is not the same
quantity as in JJ models.

In the central regions both the integral in \cref{eq:rhosigma2p} and $\Sigma_*(R)$ 
diverge, so that $\sigp(R)$ can be properly defined
only as a limit. For what concerns the {\it galaxy} contribution we have
\begin{equation}
\sigp(0)=\srad(0), \qquad \sa >0,
\label{eq:sigp0=srad0}
\end{equation}
where $\srad(0)$ is given by \cref{eq:srad2_0}. For $\ra =0$, instead,
$\sigp^2(R) \sim -\,\srad^2(0)\ln\eta$. In presence of the central BH,
$\srad(r)$ and $\sigp(R)$ are dominated by the BH contribution, and from
eqs. (\ref{eq:BH_dom}) and (\ref{eq:rhosigma2p}) we obtain
\begin{equation}
\sigp^2(R)\sim\frac{2\Psin\mu}{3\pi\eta}, \qquad \sa\geq 0.
\label{eq:sigp_BH}
\end{equation}
The independence of $\sigp(R)$ from the specific value of $\sa$ in the
central regions is shown in the bottom panel of
Fig. \ref{fig:sigma_3D_p}. Note that eqs. (\ref{eq:sigp0=srad0}) and (\ref{eq:sigp_BH})
coincide with their analogues for JJ models.

\section{Virial, potential, and  kinetic energies}\label{sec:Virial}

Among the several global quantities associated with a stellar system,
those entering the Virial Theorem (hereafter, VT) are certainly the
most interesting for many observational and theoretical studies (e.g.,
Ciotti 2000, Binney \& Tremaine 2008). For the stellar component of J3
models we have
\begin{equation}
2\Ks = -\,\Ws \equiv -\,(\Wg+\Wbh),
\label{eq:Ks}
\end{equation}
where 
\begin{equation}
\Ks=2\pi\int_0^{\infty}\!\rhos(r)\!\left[\srad^2(r)+\stan^2(r)\right]\!r^2dr \equiv K_{*\rm{g}}+K_{\rm *BH}
\label{eq:Kstar}
\end{equation} 
is the total kinetic energy of the stars,
\begin{equation}
\Wg=-\,4\pi G\!\int_0^{\infty}\!\rhos (r)\Mg(r)rdr
\end{equation}
is the interaction energy of the stars with the gravitational field of the galaxy (stars plus DM), and finally
\begin{equation}
\Wbh=-\,4\pi G \Mbh \int_0^\infty\!\rhos (r)rdr
\end{equation}
is the interaction energy of the stars with the central BH. For a
Jaffe galaxy $\Wbh$ diverges; the VT implies that also $K_{\rm *BH}$
diverges.

\begin{figure}
 \centering
  \includegraphics[scale=1.11]{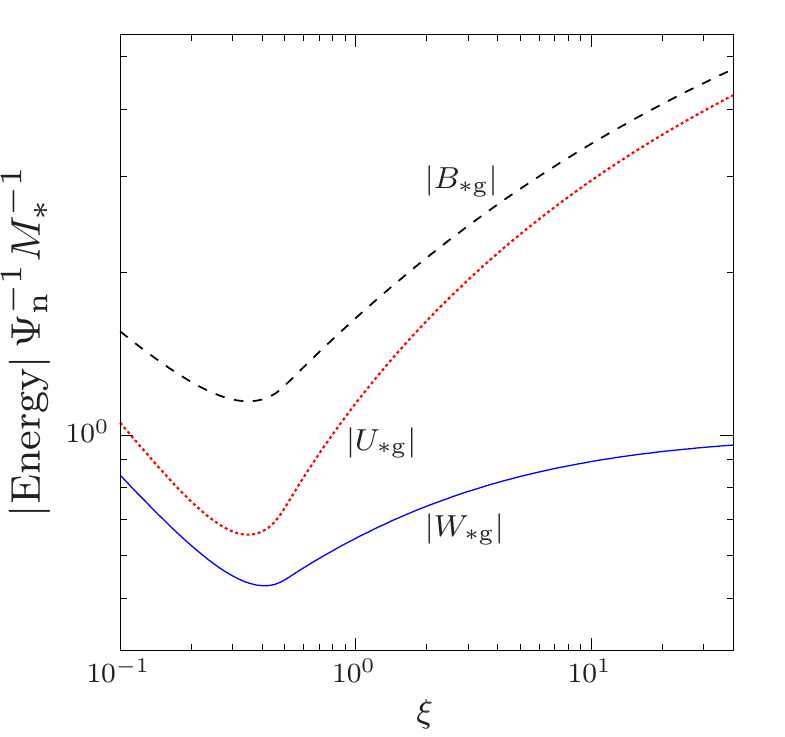}
 \vspace*{-5mm}
 \caption{Absolute values of the normalized gravitational energies for
   the stellar component, as a function of $\xi$, in case of minimum
   halo and absence of a central BH. See Sect. \ref{sec:Virial} for
   more details.}
\label{fig:virial}
\end{figure}

The contribution of the total galaxy potential to $\Wg =\Wss +\Wdm$
(where $\Wss$ is due to the self-interaction of the stellar
distribution, and $\Wdm$ to the effect of the DM halo) is finite, and
$\Wg$ is written as
\begin{equation}
\Wg=-\,\Psin\Ms\MR\,\times\cases{
            \displaystyle{\Hcsi(\xi,0)-\frac{\ln\xi}{\xi - 1}},
                       \qquad\, \xi\ne1;
         \cr\cr
         \displaystyle{\Hcsi(1,0)-1}\,,
                                                     \quad\quad\qquad \xi=1,
         }
\label{eq:Wg}
\end{equation}
where the function $\Hcsi(\xi,s)$ is given in Appendix \ref{App_C}. 
More generally, $\Wg$ is a finite quantity for the stellar component
of J3 models; it follows that it is possible to define the
(3-dimensional) stellar virial velocity dispersion as $\sigma^2_{\rm
  V}=-\,\Wg/\Ms$. Once $\Wss$ is known, $\Wdm$ is immediately obtained
as $\Wg-\Wss$, and we have
\begin{equation}
\Wss=-\,4\pi G\!\int_0^{\infty}\!\rhos (r)\Ms(r)rdr=-\,\frac{\Psin\Ms}{2}.
\end{equation}
As well known, in multi component systems the energy $W$ of a given
component is {\it not} the potential energy of the component itself in
the total potential. Therefore, we shall now calculate the different
contributions to the potential energy $\Us$ of the stellar
component. As done for the interaction energy $\Ws$, we can decompose
the potential energy $\Us$ as
\begin{equation}
\Us=\,\Ug\,+\,\,\Ubh,
\end{equation}
where 
\begin{equation}
\Ug =\,\Uss+\,\Udm={1\over 2}\int\!\rhos\Phis\,d^3\mathbf{x}\,+\!\int\!\rhos\Phidm\,d^3\mathbf{x}.
\label{eq:Ug}
\end{equation}
As well known,
\begin{equation}
\Ubh=\int\!\rhos \Phibh\,d^3\mathbf{x}=\Wbh,
\end{equation}
and so $\Ubh$ diverges as $\Wbh$. Moreover, the self-gravitational
energy and the virial self energy of each density component coincide,
and in our case $\Uss=\Wss$. In order to derive $\Ug$ and $\Udm$ in
\cref{eq:Ug}, we calculate the quantity
\begin{equation}
\Bg \equiv\int\!\rhos \Phig\,d^3\mathbf{x}=-\,\Psin\Ms\MR\,\Hcsi(\xi,0).
\end{equation}
With this definition one has that $\Ug=\Bg-\Uss$, and
$\Udm=\Bg-\,2\Uss$. The trends of $|\Ug|$, $|\Bg|$, and $|\Wg|$, as
function of $\xi$, are shown in Fig. \ref{fig:virial}.

Note that $\Bg$ {\it is not} the gravitational energy $\Ug$ of the
stars in the galaxy total potential, but an important quantity in the theory of
galactic winds, and in studies of the hot gas content of early-type
galaxies. For examples, it measures the energy per unit time
($L_{\rm{grav}}$) to be provided to the ISM of early-type galaxies
(via, e.g., supernova explosions, or thermalization of the velocity of
stellar winds, or AGN feedback) in order to steadily extract the mass
input injected over the galaxy body from evolving stars. This power is
given as $L_{\rm{grav}}\propto|\Bg|$ (e.g., see Pellegrini 2011,
Posacki et al. 2013).

\subsection{Stability}\label{sec:stability}

A particularly relevant application of the VT is in the determination
of the conditions required to prevent the onset of the Radial Orbit
Instability (hereafter, ROI). Indeed, stellar systems supported by a
large amount of radial orbits are in general unstable (e.g., Fridman
\& Polyachenko 1984, and references therein). As shown by several
numerical studies of one and two component systems (see, e.g., Merritt
\& Aguilar 1985; Bertin \& Stiavelli 1989; Saha 1991, 1992; Bertin et
al. 1994; Meza \& Zamorano 1997; Nipoti, Londrillo \& Ciotti 2002),
quantitative information about stability can be obtained by using the
function
\begin{equation}
\Xi \equiv {2 \Krad\over\Ktan} = -\,{4\over 2+\Wg/\Krad}.
\label{eq:Xi}
\end{equation}
where $\Krad$ and $\Ktan =\Ks -\Krad$ are the total
kinetic energy of stars, associated with the radial and tangential
components of the velocity dispersion tensor, respectively (see
eq. [\ref{eq:Kstar}]), and the last expression in \cref{eq:Xi} derives
from \cref{eq:Ks}. Of course, we {\it exclude} the effect of the
central BH, to avoid the divergence of the kinetic energy
$K_{\rm *BH}$. From its definition, $\Xi\to1$ for $\sa\to\infty$
(globally isotropic models), whereas $\Xi\to\infty$ for $\sa\to 0$
(fully radially anisotropic models). Note that, for $\mu=0$, $\Xi$ is
independent of $\MR$. Here we adopt the usual empirical requirement 
for stability of $\Xi<1.70 \pm 0.25$.

With an analogous treatment to that adopted to derive \cref{eq:kernel_K}, it can be shown that 
\begin{equation}
\Krad\,=\,2\pi G\!\int_0^\infty\!{\cal L}(r)\rhos(r)\Mg(r)\!\left(1+\frac{\ra^2}{r^2}\right)\!dr,
\end{equation}
where
\begin{equation}
{\cal L}(r)=r-\ra\,\atan\,\frac{r}{\ra},
\end{equation}
which tends to $r$ in the limit $\ra \to 0$, in agreement with the VT in \cref{eq:Ks}.

Unfortunately $\Krad$ cannot be expressed via elementary functions, so
that we explore numerically the fiducial stability condition
$\Xi(\sa,\xi)=1.7$. In Fig. \ref{fig:wsc} with the green dashed curve
we plot the resulting lower bound for stability $\sa(\xi)$. The
resulting trend, i.e. the fact that $\sa(\xi)$ increases with $\xi$,
is in agreement with the behaviour of other families of one and two
component models (Ciotti 1996, 1999, CMZ09, CZ18, see also Carollo et
al. 1995). As already discussed in CZ18, in order to guarantee the
stability of the system, the increase of $\sa$ with $\xi$ is a simple
consequence of the orbital distribution in OM models, which are
radially anisotropic for large $r$ and isotropic in the central
region. Moreover, $\sa(\xi)$ is quite larger than the critical values
$\sa$ for consistency, so that the maximally radially anisotropic
models with positive DF are almost certainly unstable.

\section{Conclusions}\label{sec:Conclusions}

A new family of spherical, two-component galaxy models is presented,
following the approach introduced in CMZ09 and CZ18. These models,
called J3 models, have a stellar component described by the Jaffe
density profile, and a {\it total} density component such that the
resulting DM halo -- defined as the difference between the total and
the stellar density distributions -- can be made asymptotically
identical to a NFW profile, both at the center and at large
radii. This property makes the J3 models an improvement over JJ
models, while retaining the same analytical simplicity. A BH is also
added at the center of the system, and the orbital structure of the
stellar component follows the Osipkov-Merritt anisotropy profile. The
models are fully determined once the total stellar mass ($\Ms$) and
its scale length ($\rs$) are assigned, together with a
total-to-stellar density ratio ($\MR$), a total-to-stellar scale
length ratio ($\xi$), a BH-to-stellar mass ratio ($\mu$), and finally
the anisotropy radius ($\ra$) of the stellar component. The J3 models
allow for an almost complete analytical treatment with quite simple
explicit expressions of several quantities of interest in
observational and theoretical works.  The main results can be
summarized as follows.

\begin{itemize}

\item We derive analytical constraints on $\MR$ and $\xi$ to assure
  positivity and monotonicity of the DM halo density distribution. For
  a given $\xi$, the model corresponding to the minimum value allowed
  for $\MR$ is called {\it minimum halo model}. In particular, for
  $\xi\geq 1/2$ the positivity and monotonicity conditions coincide, requiring
  $\MR\geq\xi$. For arbitrary choices of $\MR$ and $\xi$, near the
  origin the DM density profile diverges as $\rhoDM(r)\propto r^{-2}$,
  but in the minimum halo case with $\xi>1/2$ the models are centrally
  ``baryon dominated'', with $\rhoDM(r)\propto r^{-1}$, as for the NFW
  profile. Moreover, at large radii $\rhoDM(r)$ is, by construction,
  always proportional to $r^{-3}$, again following the NFW
  profile. Two simple formulae determine the parameters of the NFW
  model {\it identical} to $\rhoDM(r)$ at the center and at large
  radii; remarkably, the NFW profile so obtained stays close to
  $\rhoDM(r)$ also in the intermediate region.

\item The minimum value of the OM anisotropy radius $\ra$,
  corresponding to a dynamically consistent stellar component, is
  first estimated using the necessary and sufficient conditions given
  in CP92. The consistency analysis is then performed for the
  isotropic DM halo and in presence of a central BH; it is found that
  for $\xi\geq 1$, once {\it positivity only} of $\rhoDM(r)$ is
  assured, i.e. $\MR\geq\xi$, $\rhoDM$ automatically satisfies the NC
  and WSC, and so it can be supported by a nowhere negative
  phase-space DF.

\item The DF is then recovered numerically, and we illustrate its
  behavior for a few representative cases, for different choices of
  $\MR$, $\xi$, and $\ra$. Then we determined the (minimum) critical
  value of $\ra$, as a function of the model parameters, finding a
  curve that nicely parallels that given by the WSC.  We showed that
  in absence of the central BH, the minimum value of $\ra$ depends
  only on $\xi$, and it is independent of $\MR$.  For example, for
  models with $\xi=13$ and no BH, the positivity of the DF requires
  $\ra\gtrsim 0.0143\,\rs$.  In particular, $\ra$ decreases for
  increasing $\xi$, i.e., a DM halo more extended than the stellar
  distribution increases the ability of the stellar component to
  sustain radial anisotropy. On the contrary, more concentrated DM
  halos require a more isotropic orbital distribution. This behavior
  is similar to what already found for JJ models.

\item Having determined the region of the parameter space
  corresponding to physically consistent models, we solved
  analytically the Jeans equations for the stellar component, for
  generic values of the model parameters. The asymptotic expansions of
  $\srad(r)$ and $\sigp(R)$ near the center and at large radii were
  obtained; it is shown that J3 models, in the central region, behave
  identically to JJ models. Accordingly when $\mu=0$ and for all
  values of $\ra >0$ (isotropic case included),
  $\sigp^2(0)=\srad^2(0)=\Psin\MR/(2\xi)$. In presence of the BH, in
  the central regions $\srad^2(r)\propto r^{-1}$, and $\sigp^2(R)\sim
  2\Psin\mu\rs/(3\pi R)$, independently of $\ra$. These results can be
  used, among other applications, to estimate the size of the
  so-called ``sphere of influence'' of the central BH.

\item Finally, the analytical expressions for the quantities entering
  the Virial Theorem, such as the stellar kinetic energy, the
  interaction energy, and the potential energies, are derived as a
  function of the model parameters. We also evaluated numerically the
  minimum value of $\ra$ corresponding to the fiducial value of
  $\simeq 1.7$ for the Friedmann-Poliachenko-Shuckman instability
  indicator, so that more anisotropic models are prone to the onset of
  Radial Orbit Instability. The minimum $\ra$ for stability increases
  for increasing $\xi$, and (in absence of the central BH) its value
  depends only on $\xi$, being independent of $\MR$.
\end{itemize}

We conclude by noting that J3 models can be a useful
starting point for more advanced modeling of the dynamics of
elliptical galaxies, and can be easily implemented in numerical
simulations. In addition, it can be shown that J3 models allow for a
fully analytical treatment of Bondi accretion along the lines
discussed elsewhere (Ciotti \& Pellegrini 2017, 2018); we defer this
study to a future work.

\section*{Acknowledgments}
We thank the anonymous referee for useful comments, that improved the
presentation.




\appendix

\section{Projected densities}\label{App_A}

For the stellar density, the functions $\fstar(\eta)$ and $\gs(\eta)$
appearing in eqs. (\ref{eq:sigs(R)}) and (\ref{eq:proj_stellar_mass}) are given by
\begin{equation}
\fstar(\eta)=
\cases{
            \displaystyle{
{1\over 4\eta} +
{\sqrt{1-\eta^2}-(2-\eta^2)\,{\rm arcsech\,\eta}\over 2\pi (1-\eta^2)^{3/2}}},
                                                     \quad 0<\eta <1,
         \cr\cr
         \displaystyle{{1\over 4} - {2\over 3\pi}},
                       \qquad\qquad\qquad\qquad\qquad\qquad\quad\quad\, \eta=1,
         \cr\cr
         \displaystyle{
{1\over 4\eta}-
{\sqrt{\eta^2-1}+(\eta^2-2)\,{\rm arcsec\,\eta}\over 2\pi(\eta^2-1)^{3/2}}},
                      \qquad\,\,\, \eta >1,}
\label{eq:fstar}
\end{equation}
and
\begin{equation}
\gs(\eta)=\eta \times \cases{
            \displaystyle{{\pi\over 2}-\eta\,\frac{{\rm arcsech}\,\eta}{\sqrt{1-\eta^2}}},
                                                     \quad\quad 0<\eta <1,
         \cr\cr
         \displaystyle{{\pi\over 2}-1},
                       \qquad\qquad\qquad\quad\,\, \eta=1,
         \cr\cr
         \displaystyle{{\pi\over 2}-\eta\,\frac{{\rm arcsec}\,\eta}{\sqrt{\eta^2-1}}}\,,
                      \qquad\quad \eta >1,}
\label{eq:gstar}
\end{equation}
where $\eta=R/\rs$. For the total galaxy density, the analogous
functions $\fgal(\eta)$ and $\ggal(\eta)$
appearing in eqs. (\ref{eq:Sigg(R)}) and (\ref{eq:proj_total_mass}) read
\begin{equation}
\fgal(\eta)=\cases{
            \displaystyle{\frac{1}{4\eta}-\frac{{\rm arcsech}\,\eta}{2\pi\sqrt{1-\eta^2}}},
                                                     \qquad 0<\eta <1,
         \cr\cr
         \displaystyle{\frac{1}{4}-\frac{1}{2\pi}},
                       \qquad\qquad\qquad\quad\,\, \eta=1,
         \cr\cr
         \displaystyle{\frac{1}{4\eta}-\frac{{\rm arcsec}\,\eta}{2\pi\sqrt{\eta^2-1}}},
                      \qquad\quad \eta >1,}
\label{eq:fgal}
\end{equation}
and
\begin{equation}
\ggal(\eta)=\eta\times\cases{
            \displaystyle{\frac{\pi}{2}-\frac{\ln(2/\eta)}{\eta}+\frac{\sqrt{1-\eta^2}\,{\rm arcsech}\,\eta}{\eta}},
                                                     \quad 0<\eta <1,
         \cr\cr
         \displaystyle{\frac{\pi}{2}-\ln 2},
                       \qquad\qquad\qquad\qquad\qquad\quad\quad\quad\,\, \eta=1,
         \cr\cr
         \displaystyle{\frac{\pi}{2}-\frac{\ln(2/\eta)}{\eta}-\frac{\sqrt{\eta^2-1}\,{\rm arcsec}\,\eta}{\eta}},
                      \qquad\,\,\, \eta >1,}
\label{eq:ggal}
\end{equation}
where now $\eta=R/\rg$.

\section{Positivity and monotonicity of the dark matter halo}\label{App_B}

The condition for the {\it positivity} of the DM halo density is obtained by imposing $\rhoDM(r)\geq 0$, i.e.
\begin{equation}
\MR\,\ge\,\frac{\xi + s}{(1+s)^2}\,, \quad\quad s\ge 0\,.
\end{equation}
Therefore, in order to have a nowhere negative DM halo for given
$\xi$, $\MR$ must be greater than or equal to the maximum $\Rm(\xi)$
of the function above, and simple algebra proves \cref{eq:pos}.

The {\it monotonicity} condition for $\rhoDM$ is then obtained by requiring that $d\rhoDM(r)/dr\leq 0$, i.e.
\begin{equation}
\MR\,\geq\,\frac{2 (\xi + s)^2 (1+2s)}{(1+s)^3 (2\xi + 3s)}\equiv\Mmon(\xi,s)\,,\quad s \ge 0\,,
\label{eq:mon}
\end{equation}
and by the same argument as above, we call $\Rmon(\xi)$ the maximum of the function $\Mmon(\xi,s)$. We obtain 
\begin{equation}
\frac{d\Mmon(\xi,s)}{ds}\propto (1-2\xi)[\xi+(3+4\xi)s]+6(1-3\xi)s^2-6s^3,
\label{eq:Mmon}
\end{equation}
where proportionality indicates product by a strictly positive
function. The explicit solution of \cref{eq:Mmon} presents no
difficulty, however important information can be easily obtained
without solving it. First, for $\xi\geq 1/2$ \cref{eq:Mmon} is nowhere
positive, and so the maximum of $\Mmon$ is reached at the center, with
$\Rmon(\xi)\equiv\Mmon(\xi,0)=\xi=\Rm(\xi)$. In practice, {\it for
  $\xi\geq 1/2$ the positivity and monotonicity conditions coincide},
similarly to the case of JJ models, where the positivity and
monotonicity conditions coincide for all values of $\xi$. For
$0<\xi<1/2$ the situation is different. In fact, the cubic function in
\cref{eq:Mmon} is positive for small values of $s$, and negative for
large values of $s$, and so $d\Mmon(\xi,s)/ds$ has at least one zero
(and $\Mmon$ at least one maximum) for $s\geq 0$. Moreover, the
Descartes Theorem shows that for $0<\xi < 1/2$ at most one positive
zero of the cubic exists, therefore corresponding to the single
maximum $\Rmon(\xi)$: as we are mainly interested in realistic models
with $\xi\geq 1$, we do not discuss further this case.

As illustrated in Sect. \ref{sec:NC_SC}, for an isotropic DM halo the
NC is equivalent to the monotonicity of the density profile, and from
the results obtained above, for $\xi\geq 1/2$ positivity, monotonicity
and NC coincide, and hold for $\MR\geq\xi$. The WSC for the isotropic
halo in the total galactic potential and in {\it absence} of the
central BH involves a cumbersome function, and so we do not give here
its expression: a numerical study indicates however that for $\xi\geq
1$ the WSC holds when $\MR\geq\xi$. The WSC for the isotropic halo in
the potential of a dominant central mass reduces to
\begin{equation}
\MR\,\geq\,\frac{(\xi+s)^3(6s^2+4s+1)}{(1+s)^4(3s^2+3\xi s+\xi^2)},
\label{eq:wsc_halo_BHdominant}
\end{equation}
and simple algebra shows that for $\xi\geq 1$ the maximum is reached
at $s=0$, i.e. again for $\MR\geq\xi$. Therefore, from the discussion
after \cref{eq:WSC} we conclude that for $\xi\geq 1$ and $\MR\geq\xi$ the
WSC for the isotropic DM halo with a central BH of arbitrary mass
holds.

As for the halo, also for the Jaffe stellar distribution the WSC leads
to an analytical but cumbersome expression. In absence of the central
BH ($\mu=0$) the WSC reduces to
\begin{equation}
\sa^2\geq-\,\frac{s^3\!\left[(\xi+s)(s-2)\!\ln(1+s/\xi)+s(1+s)\right]}{(\xi+s)(6s^2+4s+1)\!\ln(1+s/\xi)+s(1+s)(1+2s)},
\label{eq:wsc_stellar}
\end{equation}
for $s\geq 0$, where it can be noticed that the parameter $\MR$ does
not appear. The result of the numerical investigation of the
inequality above is represented by the red dotted line in
Fig. \ref{fig:wsc}, where we show the minimum value $\sam(\xi)$.

\section{The function $\Hcsi(\lowercase{\xi,s})$}\label{App_C}

The function $\Hcsi$ is defined as 
\[
\Hcsi(\xi,s)\,\equiv\int_s^\infty \ln\!\left(1+\frac{1}{t}\right)\frac{dt}{\xi + t}\,=\,\Litwo\!\left(\frac{1}{1+s}\right)\,+
\]
\begin{equation}
\cases{
      \displaystyle{\frac{1}{2}\ln\frac{1+s}{\xi+s}\ln\frac{(1-\xi)^2(1+s)}{\xi+s} + \Litwo\!\left[\frac{\xi(1+s)}{\xi+s}\right]-\Litwo(\xi),}\,
      \cr\cr
      \displaystyle{0,}\,
      \cr\cr
      \displaystyle{\!\ln\frac{\xi}{\xi-1}\ln\frac{\xi+s}{1+s} - \Litwo\!\left[\frac{\xi+s}{\xi(1+s)}\right] + \Litwo\!\left(\frac{1}{\xi}\right)\!,}\,
      }
\label{eq:Hcsi}
\end{equation}
where the three expressions (that can be obtained with some work
starting from the change of variable $1+t= 1/y$) refer to
$0<\xi<1$, $\xi=1$, and $\xi>1$, respectively, and
\begin{equation}
\Litwo(x)=-\int_0^x\frac{\ln(1-t)}{t}\,dt,\quad x\leq 1,
\end{equation}
is the dilogarithm function (e.g., Gradshteyn $\&$ Ryzhik 2007, see
also Lewin 1981), with the Euler identity $\Litwo(1)=\pi^2/6$.  At the
center,
\begin{equation}
\Hcsi(\xi,0)=\cases{
      \displaystyle{\frac{\pi^2}{3}\!+\frac{(\ln\xi)^2}{2} -\ln\xi\,\ln (1-\xi) - \Litwo(\xi),}
      \cr\cr
      \displaystyle{\frac{\pi^2}{6}}, 
      \cr\cr\
      \displaystyle{(\ln\xi)^2 -\ln\xi\,\ln (\xi -1) +\Litwo\!\left(\frac{1}{\xi}\right),}
      }
\end{equation}
while a direct expansion of \cref{eq:Hcsi} proves that 
\begin{equation}
\Hcsi(\xi,s)=
\cases{
\displaystyle{\Hcsi(\xi,0)+{s\ln s\over \xi} +
  \mathcal{O}\!\left(s\right)\!,}\cr\cr
\displaystyle{
{1\over s}-{2\xi +1\over 4s^2}+ \mathcal{O}\!\left({1\over s^3}\right)\!,}
}
\end{equation}
for $s\to 0$ and $s\to\infty$, respectively.


\end{document}